\documentclass[12pt]{article}
\usepackage{epsfig}
\usepackage{a4,isolatin1}
\usepackage{oldgerm}
\usepackage{amsthm}
\usepackage{amsmath,mathrsfs,amssymb,amscd}
\usepackage{cite}

\newlength{\dinwidth}
\newlength{\dinmargin}
\newtheorem{theorem}{Theorem}[section]
\newtheorem{proposition}[theorem]{Proposition}
\newtheorem{lemma}[theorem]{Lemma}

\newcommand{\Ibb}[1]{ {\rm I\ifmmode\mkern -3.6mu\else\kern -.2em\fi#1}}
\newcommand{\ibb}[1]{\leavevmode\hbox{\kern.3em\vrule
     height 1.2ex depth -.3ex width .2pt\kern-.3em\rm#1}}

\def\ie{{\it i.e.\,}}

\begin{document}
\bibliographystyle{plain}

\title{Modular Nuclearity and Localization}
\author{Detlev Buchholz and Gandalf Lechner\\Institut für
  Theoretische Physik, Universität Göttingen\\37077
  Göttingen, Germany
\date{}
}

\maketitle
\abstract{ 
\noindent Within the algebraic setting of quantum field 
theory, a condition is given which implies that the 
intersection of algebras generated by field operators 
localized in wedge--shaped regions of
two--dimensional Minkowski space is non--trivial; 
in particular, there exist compactly localized 
operators in such theories which can be interpreted as
local observables. The condition is based on 
spectral (nuclearity) properties of the modular 
operators affiliated with wedge algebras and the vacuum
state and is of interest in the algebraic approach 
to the formfactor program, initiated by Schroer. It is
illustrated here in a simple class of examples.
}

\section{Introduction}

There is growing evidence that algebraic 
quantum field theory \cite{Ha} not only is useful in  
structural analysis but provides also a framework for
the construction of models. Basic ingredients in 
this context are, on the one hand, the algebras affiliated 
with wedge shaped regions in Minkowski space, called
wedge algebras for short. On the other hand there 
enter the modular groups corresponding to these 
algebras and the vacuum state by Tomita--Takesaki theory.

The wedge algebras are distinguished by the fact that
the associated modular groups can be interpreted as 
unitary representations of specific Poincar\'e 
transformations. This fact was established first by Bisognano 
and Wichmann in the Wightman framework of quantum field theory 
\cite{BiWi} and, more recently, by Borchers in the algebraic 
setting \cite{Bo3}, cf.\ also \cite{Fl,Mu2}. It 
triggered attempts to construct families of such algebras 
directly within the algebraic framework \cite{BrGuLo,Mu1}.

A particularly interesting development was   
initiated by Schroer \cite{Sch} who, starting 
from a given factorizing scattering matrix in two 
spacetime dimensions, recognized how one may reconstruct 
from these data a family of wedge algebras satisfying 
locality. A complete construction of these algebras for a simple class
of scattering matrices was given in \cite{Le}. These results 
are a first important step in an algebraic 
approach to the formfactor program,  
\ie the reconstruction of quantum fields from a scattering
matrix \cite{BeKaWe,KaWe,Sm}; for more recent 
progress on this issue see also \cite{BaFrKaZa,BaKa,CaFr}. 

The second step in this approach consists in showing that,
besides field operators localized in wedges, there appear also 
local observables, \ie opera\-tors which are localized in compact 
spacetime regions, such as double cones. As any double cone in 
two dimensions is the intersection of two opposite wedges,   
local observables ought to be elements of the intersection 
of wedge algebras. The question of whether these 
intersections are non--trivial turned out to be a difficult 
one, however, and has not yet been settled. Some 
ideas as to how this problem may be tackled in 
models are discussed in \cite{SchWi}.

It is the aim of the present letter to point out an alternative
strategy for the proof of the non--triviality of the intersections 
of wedge algebras. By combining results scattered in the 
literature and casting them into a simple condition, we will 
show that the non--triviality of these intersections can be 
deduced from spectral (nuclearity) properties of the modular 
operators on certain specific subspaces of the Hilbert space. 
Thus the algebraic problem of determining intersections of
wedge algebras amounts to a problem in spectral analysis 
which seems to be better tractable. 

The subsequent section contains an abstract version of our nuclearity
condition and a discussion of its consequences in a
general algebraic setting. In Section 3 these results 
are carried over to a family of theories with factorizing S--matrix
in two--dimensional Minkowski space.
It is shown that compactly localized operators exist in any
theory complying with our condition. 
Section 4 illustrates the type of computations needed 
to verify this condition in a simple example. The 
article closes with a brief outlook.

\section{\hspace*{-5mm} Modular \!\!\! nuclearity \!\!\! and \!\!\! its \!\!\! 
consequences}
\setcounter{equation}{0}

In this section we present our nuclearity condition 
in a general setting, extracted from the more concrete structures in 
field theoretic models, and discuss its implications. 
We begin by introducing our notation and listing our
assumptions. \\[1mm]
(a) Let $\cal H$ be a Hilbert space and let $U$ 
be a continuous unitary representation of ${\mathbb R}^2$
acting on $\cal H$. Choosing proper coordinates on ${\mathbb R}^2$, 
$x = (x_0,x_1)$, the joint spectrum of the 
corresponding generators $(P_0,P_1)$ of $U$ is contained in the cone 
$V_+ \doteq \{ (p_0,p_1) \in {\mathbb R}^2: p_0 \geq |p_1| \}$ 
and there is an (up to 
a phase unique) unit vector $\Omega \in \cal H$ which is 
invariant under the action of $U$.  \\[1mm]
(b) There is a von Neumann algebra $\cal M \subset B(H)$
such that for each element $ x $ of the wedge  
$ W \doteq \{ y \in {\mathbb R}^2 : |y_0| + y_1 < 0 \}$
the adjoint action of the unitaries $U(x)$
induces endomorphisms of $\cal M $,
\begin{equation} \label{inclusion}
{\cal M} (x) \doteq U(x) {\cal M} U(x)^{-1} \subset {\cal M},
\quad x \in W.
\end{equation}
Moreover, $\Omega$ is cyclic and separating for $\cal M $. \\[1mm]
\indent It is well known that, under these circumstances, 
the algebraic properties of $\cal M $ are strongly restricted.
As a matter of fact, disregarding the trivial possibility 
that $\cal H$ is one--dimensional and $\cal M = {\mathbb C}$,
the following result has been established in  
\cite[Thm.\ 3]{Lo}. 
\begin{lemma} Under the preceding two conditions the algebra
$\cal M $ is a factor of type III$_1$ according to the classification
of Connes.
\end{lemma}
It immediately follows from this result that the algebras 
${\cal M} (x)$ are factors of type III$_1$ as well. Little is 
known, however, about the algebraic structure of the 
relative commutants ${\cal M} (x)^\prime \cap {\cal M} $ of 
${\cal M} (x)$ in ${\cal M}$, $x \in W$. Even the question
of whether these relative commutants are non--trivial has not
been settled in this general setting. Yet this question 
turns out to have an affirmative answer and, as a 
matter of fact, the algebraic structures
are completely fixed if the inclusions (\ref{inclusion}) 
are split, \ie if for each $x \in W$ there is a factor 
$\cal N$ of type I$_\infty$ such that 
\begin{equation} \label{split}
{\cal M} (x) \subset {\cal N} \subset {\cal M}.     
\end{equation}

First, the split property implies that ${\cal M}$ is
isomorphic to the unique \cite{Hag} hyperfinite
factor of type III$_1$. We briefly recall here the
argument: As $\Omega$ is cyclic and separating for 
${\cal M}$, and hence for ${\cal M} (x)$, this is also 
true for $ {\cal N}$. It follows that $ {\cal N}$,
being of type I$_\infty$, is separable in the ultraweak 
topology and consequently $\cal H$
is separable, cf.\ \cite[Prop.\ 1.2]{DoLo}. Now, 
as $U$ is continuous, ${\cal M}$ is continuous from 
the inside, ${\cal M} = \bigvee_{x \in W} {\cal M} (x)$. 
The split property thus implies that  ${\cal M}$ 
can be approximated from the inside by 
separable type I$_\infty$ factors and therefore is 
hyperfinite, cf.\ \cite[Prop.\ 3.1]{BuDaFr}.
Knowing also that it is of type III$_1$, the
assertion follows.

Secondly, the split property implies that 
${\cal M} (x)^\prime \cap {\cal M} $, $x \in W$,
is isomorphic to the hyperfinite factor of type III$_1$ as well. 
This can be seen as follows \cite{DoLo}. On a separable 
Hilbert space $\cal H$, any factor of type III has cyclic and
separating vectors \cite[Cor.\ 2.9.28]{Sa}.
Moreover, for {\it any}  von Neumann algebra on $\cal H$ with 
a cyclic and a separating vector there exists a dense $G_\delta$
set of vectors which are both, cyclic and 
separating \cite{DiMa}.
Now, taking into account that  ${\cal N}$ is isomorphic
to $\cal B(H)$, the relative commutant ${\cal M} (x)^\prime \cap {\cal N}$
of the type III factor 
${\cal M} (x)$ in ${\cal N}$ is 
(anti)isomorphic to  ${\cal M} (x)$ by Tomita--Takesaki theory.
It is therefore of type III and has cyclic vectors in $\cal H$.
This holds {\it a fortiori} for 
${\cal M} (x)^\prime \cap {\cal M} \supset {\cal M} (x)^\prime \cap {\cal N}$
and, as $\Omega$ is separating for ${\cal M}$, 
the relative commutant ${\cal M} (x)^\prime \cap {\cal M}$
has a dense  $G_\delta$ set of cyclic and separating vectors.
But the intersection of a finite number of dense $G_\delta$ sets
is non--empty. So we conclude that the triple ${\cal M}$,
${\cal M} (x)$ and ${\cal M} (x)^\prime \cap {\cal M}$ has a joint
cyclic and separating vector in $\cal H$. The inclusion
(\ref{split}) is thus a standard split inclusion according
to the terminology in \cite{DoLo}. In particular, there is 
a spatial isomorphism mapping ${\cal M} (x) \bigvee {\cal M}^\prime$
on $\cal H$ onto ${\cal M} (x) \otimes {\cal M}^\prime$ on 
$\cal H \otimes \cal H$ \cite{DaLo}. By taking commutants, 
we conclude that ${\cal M} (x)^\prime \cap {\cal M} $
is isomorphic to ${\cal M} (x)^\prime \otimes {\cal M}$, $x \in W$.
The statement about the algebraic structure of the relative
commutant then follows.

It seems difficult, however, to establish the existence of
intermediate type I$_\infty$ factors $\cal N$ in the inclusions 
(\ref{split}) for concretely given $\{{\cal M},U,{\cal H} \}$, 
and this may be the reason why this strategy of establishing 
the non--triviality of relative commutants has been 
discarded in \cite{SchWi}. Yet the situation is actually not  
hopeless, the interesting point being that the existence of 
the desired factors can be derived from spectral properties of the modular 
operator $\Delta$ affiliated with the pair $({\cal M}, \Omega)$.
Recalling that a linear map from a Banach space
into another one is said to be nuclear if it can be decomposed 
into a series of maps of rank one whose norms are summable, we extract
the following pertinent condition  
from \cite{BuDaLo1}. \\[1mm]
(c) Modular Nuclearity Condition: For any given $x \in W$ the map 
\begin{equation} \label{nuclearity}
M \mapsto \Delta^{1/4} M \Omega, \quad M \in {\cal M} (x), 
\end{equation}
is nuclear. Equivalently, 
since $\Delta^{1/4}$ is invertible, the
image of the unit ball in  ${\cal M} (x)$ under this map is a nuclear subset 
of $\cal H$. \\[1mm]
\indent Since $\Omega$ is cyclic and separating for 
$\cal M$ and the algebras ${\cal M} (x)$, $\cal M$ both are factors, 
it follows from the modular nuclearity condition (c) that the inclusions
${\cal M} (x) \subset \cal M$, $x \in W$, 
are split \cite[Thm.\ 3.3]{BuDaLo1}. Conversely, if these
inclusions are split, the map (\ref{nuclearity}) has to be
compact, at least. Thus a proof of the 
split property (\ref{split}) amounts to a spectral
analysis of the operator $\Delta^{1/4}$ on the subspaces
${\cal M} (x) \, \Omega \subset \cal H$. 
This task is, as we shall see, manageable 
in concrete applications. We summarize the
results of the preceding discussion in the following proposition.
\begin{proposition} \label{prop}
Let $\{{\cal M},U,{\cal H} \}$ be a triple 
satisfying conditions (a), (b) and (c), stated above. Then,
for $x \in W$, \\[1mm]
(i) the inclusion ${\cal M} (x) \subset \cal M$ is split; \\[1mm]
(ii) the relative commutant 
${\cal M} (x)^\prime \cap {\cal M} $ is isomorphic to 
the unique hyperfinite type III$_1$ factor. In particular,
it has cyclic and separating vectors. 
\end{proposition}
We conclude this section by noting that any triple
$\{{\cal M},U,{\cal H} \}$ as 
in the preceding proposition can be used to construct a 
non--trivial Poincar\'e covariant net of local algebras on 
two--dimensional Minkowski space $\mathbb R^2$. Following closely the
discussion in \cite{Bo1,Bo3}, we first note that the 
modular group $\Delta^{is}, s \in \mathbb R$, and the modular
conjugation $J$ affiliated with $({\cal M}, \Omega)$ can 
be interpreted as representations of proper Lorentz 
transformations $\Lambda$ (having determinant one). More specifically, 
if $\Lambda$ is any such transformation and 
$\Lambda = (-1)^\sigma B(\theta)$ its polar decomposition,
where $\sigma \in \{0,1\}$ and $ B(\theta)$ is a boost with
rapidity $\theta \in \mathbb R$, one can show that
\begin{equation}
U(x,\Lambda) \doteq U(x) \, J^\sigma \, \Delta^{i\theta/{2 \pi}} 
\end{equation}
defines a continuous (anti)unitary representation of the
proper Poincar\'e group \cite{Bo1}. Moreover, $\Omega$
is invariant under the action of these operators and
may thus be interpreted as a vacuum state. Setting
${\cal R} (\Lambda W + x) \doteq U(x,\Lambda) {\cal M}
U(x,\Lambda)^{-1}$, one obtains a local (as a matter of
fact, Haag--dual) Poincar\'e covariant net of wedge algebras
on $\mathbb R^2$. Denoting  the double cones in $\mathbb R^2$
by $C_{x,y} \doteq (-W + x) \cap (W + y)$,
$x - y \in W$, the corresponding algebras
\begin{equation}
{\cal R}(C_{x,y}) \doteq {\cal R} (W + x)^\prime \cap  
{\cal R} (-W + y)^\prime
=  {\cal M} (x)^\prime \cap  {\cal M} (y)
\end{equation} 
are non--trivial according to the preceding proposition.
As was shown in \cite{Bo1}, they form a local net
on $\mathbb R^2$ which is relatively local to the wedge algebras and
transforms covariantly under the adjoint action of 
$U(x,\Lambda)$. It may thus be interpreted as a net of
local observables. The vacuum vector $\Omega$ need 
not be cyclic for the local algebras, however. 
In fact, thinking of theories exhibiting solitonic
excitations of $\Omega$ which are localized 
in wedge regions, this may also not be expected 
in general.

\section{Applications to field theoretic models}
\setcounter{equation}{0}

We carry over now the results of the preceding 
section to the framework of two--dimensional models 
and indicate their significance for the 
formfactor program, \ie the reconstruction of local 
observables and fields from a given factorizing scattering matrix.

For the sake of concreteness, we restrict attention here to
the theory of a single massive particle with 
given two--particle scattering function $S_2$, as considered
in \cite{Sch,Sch1} and described in more detail in
\cite{Le}. The Hilbert space of the theory is
conveniently represented as 
the $S_2$--symmetrized Fock space  ${\cal H} = \bigoplus_{n = 0}^\infty
{\cal H}_n$. Here the subspace ${\cal H}_0$ consists of 
multiples of the vacuum vector $\Omega$ and, 
using the parameterization of the mass shell by the rapidity $\theta$, 
\begin{equation}
p(\theta) = m \, \big(\text{ch}
(\theta), \text{sh}(\theta) \big), \quad \theta \in \mathbb R,
\end{equation}
the single particle space  ${\cal H}_1$ can be
identified with the space of square integrable functions
$\theta \mapsto \Psi_1 (\theta) $ with norm given by 
\begin{equation} 
\| \Psi_1 \|^2 = \int d\theta \, | \Psi_1 (\theta) |^2. 
\end{equation}
The elements of the $n$--particle space 
${\cal H}_n$ are represented by square integrable functions 
$\theta_1 \dots \theta_n \mapsto \Psi_n (\theta_1, \dots ,\theta_n)$
which are $S_2$--symmetric, 
\begin{equation} 
\Psi_n (\theta_1, \dots ,\theta_{i+1}, \theta_{i}, \dots, \theta_n)
= S_2(\theta_{i} - \theta_{i+1}) \, 
\Psi_n (\theta_1, \dots ,\theta_{i}, \theta_{i+1}, \dots, \theta_n).
\end{equation}
Here $\zeta \mapsto S_2 (\zeta)$ is the scattering
function which is continuous and bounded on the strip
$\{ \zeta \in {\mathbb C} : 0 \leq \text{Im} \, \zeta \leq \pi \}$, 
analytic in its interior
and satisfies, for $\theta \in \mathbb R$, the unitarity and 
crossing relations
\begin{equation} 
S_2(\theta)^{-1} = \overline{S_2(\theta)} = S_2(-\theta) =
 S_2(\theta + i \pi).
\end{equation}
On $\cal H$ there acts a continuous unitary representation $U$ of the proper 
orthochronous Poincar\'e group, given by
\begin{equation}
\big(U(x,B(\theta)) \, \Psi \big){}_n  (\theta_1, \dots ,\theta_n)
\doteq e^{i x \sum_{j=1}^n p(\theta_j)} \
 \Psi_n  (\theta_1 - \theta, \dots ,\theta_n - \theta).
\end{equation}
It satisfies the relativistic spectrum condition, \ie the 
joint spectrum of the generators $P = (P_0,P_1)$ 
of the translations $U(\mathbb R^2,1)$ is 
contained in $V_+$. Moreover, there is an antiunitary operator $J$ on
$\cal H$, representing the PCT symmetry. It is given by
\begin{equation}
\big(J \, \Psi \big){}_n   (\theta_1, \dots ,\theta_n) \doteq 
\overline{ \Psi_n  ( \theta_n, \dots , \theta_1) }.
\end{equation}

As in the case of the bosonic and fermionic Fock spaces, one can 
define creation and annihilation operators $z^\dagger (\theta)$, 
$z (\theta)$ (in the sense of operator valued distributions)
on the dense subspace ${\cal D} \subset \cal H$ of vectors 
with a finite particle number. They are hermitian conjugates with 
respect to each other and satisfy the Fadeev--Zamolodchikov relations
\begin{eqnarray}
& z^\dagger (\theta) z^\dagger (\theta^\prime) =
S_2(\theta - \theta^\prime) \,  z^\dagger (\theta^\prime) z^\dagger (\theta), 
\quad 
z (\theta) z (\theta^\prime) = 
S_2(\theta - \theta^\prime) \,  z (\theta^\prime) z (\theta), & \nonumber \\
& z (\theta) z^\dagger (\theta^\prime) =
S_2(\theta^\prime - \theta) \, z^\dagger (\theta^\prime) z (\theta)
+ \delta(\theta - \theta^\prime) \, 1. &
\end{eqnarray}
Their action on ${\cal D}$ is fixed by the equations
\begin{equation}
(z^\dagger (\theta_1) \dots z^\dagger (\theta_n) \, \Omega, \Psi) =
(n !)^{1/2} \, \Psi_n (\theta_1, \dots ,\theta_n), \quad  
z (\theta) \, \Omega = 0.
\end{equation}
With the help of these creation and annihilation operators one 
can define on ${\cal D}$ a field $\phi$, setting
\begin{equation}
\phi (f) \doteq z^\dagger (f_+) + z(f_-), \quad 
f \in {\cal S} (\mathbb R^2),
\end{equation}
where 
\begin{equation}
f_\pm (\theta) \doteq (2 \pi)^{-1} \int dx  f(x) \, e^{\pm i p(\theta) x}
\end{equation}
and we adopt the convention that, both, $z^\dagger (\, \cdot \,)$ 
and $z ( \, \cdot \,)$ are complex linear on the space of test functions.

It has been shown in \cite{Le} that $\phi$ 
transforms covariantly under the adjoint action 
of the proper orthochronous Poincar\'e group,
\begin{equation} \label{covariance}
U(x,B) \,  \phi(f) \,  U(x,B)^{-1}
= \phi(f_{x,B}),
\end{equation}
where $f_{x,B}(y) \doteq f(B^{-1}(y-x))$, $y \in \mathbb R^2$.
Moreover, $\phi$ is real, $\phi(f)^* \supset \phi(\overline{f})$, and
each vector in $\cal D$ is entire analytic for the operators $\phi(f)$.
Since $\cal D$ is stable under their action, these operators 
are essentially selfadjoint on this domain for real $f$.
We mention as an aside that the fields $\phi(f)$
are polarization--free generators in the sense of \cite{BoBuSch}.

Denoting the selfadjoint extensions of $\phi(f)$, $f$ real, by the
same symbol, one can define the von Neumann algebras
\begin{equation}
{\cal R} (W + x) \doteq \{ e^{ i \phi(f)} : \text{supp} f \subset W + x
\}^{\prime \prime}, \quad x \in {\mathbb R}^2,
\end{equation} 
where $W$ denotes, as before, the wedge 
$ W \doteq \{ y \in {\mathbb R}^2 : |y_0| + y_1 < 0 \}$. 
With the help of the PCT operator $J$ one can also define algebras 
corresponding to the opposite wedges,
\begin{equation}
{\cal R} (-W - x) \doteq J \, {\cal R} (W + x) \, J, \quad
x \in  {\mathbb R}^2.
\end{equation}

Now, given an arbitrary proper Lorentz transformation $\Lambda$
with polar decomposition $\Lambda = (-1)^\sigma B$, 
$\sigma \in \{0,1 \}$, one obtains a
representation of the proper Poincar\'e group, setting
$U(x,\Lambda) \doteq U(x,B) J^{\sigma}$.
It then follows from the covariance properties (\ref{covariance}) of the
field that
\begin{equation}
U(x,\Lambda) \, {\cal R} (\pm W + y) U(x,\Lambda)^{-1}
=  {\cal R} (\pm  (-1)^\sigma W + \Lambda y +x),
\end{equation}
taking into account that the wedge $W$ is stable under the
action of boosts. So, by this construction, one
arrives at a Poincar\'e covariant
net of wedge algebras on two--dimensional Minkowski space.

It has been shown in \cite{Le} that this net is local,
\begin{equation} \label{wedgelocality}
 {\cal R} (\pm W + x) \subset  {\cal R} (\mp W + x)^\prime,
\end{equation}
and that $\Omega$ is cyclic and separating for the wedge 
algebras (and hence for their commutants).  

The triple $\{ {\cal R} (W) , U({\mathbb R}^2,1), {\cal H} \}$
satisfies conditions (a) and (b) given in the preceding section.
More can be said by making use of modular theory
and certain specific domain properties of
the field $\phi$.
\begin{proposition} Let ${\cal R}(W)$ be the algebra 
defined above. Then \\[1mm]
(i) \, the modular group and conjugation affiliated with 
$({\cal R}(W), \Omega)$ are given by 
${\mathbb R} \ni \lambda \mapsto U(0,B(2 \pi \lambda))$ 
and $J$, respectively;  \\[1mm]
(ii)  ${\cal R}(W)^\prime = {\cal R}(-W) $ (Haag duality).
\end{proposition}
\begin{proof} Let $\Delta_W$, $J_W$ be the modular operator 
and conjugation, respectively, affiliated with 
$({\cal R}(W), \Omega)$. It follows from modular theory 
that any boost $ U(0,B)$ commutes with $\Delta_W$ and $J_W$ 
since $\Omega$ is invariant and 
${\cal R}(W)$ is stable under its (adjoint) action. 
Hence $\lambda \mapsto V(\lambda) \doteq  U(0,B(2 \pi \lambda)) 
\Delta_W^{-i\lambda}$ is a continuous unitary representation
of $\mathbb R$ with the latter properties. 
Moreover, $V(\lambda)$ commutes with all boosts 
$ U(0,B)$ and, by a theorem of Borchers \cite{Bo1}, also 
with all translations $U(x,1)$. Since the restriction of $U$ to the
proper orthochronous Poincar\'e group acts irreducibly on 
${\cal H}_1$, one concludes that $V(\lambda) \upharpoonright {\cal H}_1
= e^{i \lambda c} \, 1$ for fixed real $c$ and any
$\lambda \in \mathbb R$. 

Now, for real $f$ with $\text{supp} f \subset W$,
$\phi(f)$ is a selfadjoint operator affiliated with ${\cal R}(W)$,
and the same holds for
$\phi_\lambda(f) \doteq V(\lambda)\phi(f) V(\lambda)^{-1} $, 
$\lambda \in \mathbb R$, because of the stability of ${\cal R}(W)$
under the adjoint action of $ V(\lambda)$. So both operators commute with
all elements of  ${\cal R}(W)^\prime$. Since $\Omega$ is invariant
under the action of $V(\lambda)^{-1}$ and since 
$\phi(f) \Omega \in {\cal H}_1$, the preceding result implies
\begin{equation}
\big(\phi_\lambda(f) - e^{i \lambda c} \phi(f) \big) A^\prime \Omega = 0,
\quad A^\prime \in {\cal R}(W)^\prime.
\end{equation}
It will be shown below that the dense set of vectors 
${\cal R}(W)^\prime \Omega$ is a core, both, for $\phi(f)$
and $\phi_\lambda(f)$. Hence $\phi_\lambda(f) = e^{i \lambda c} \phi(f)$
which, in view of the selfadjointness of the field 
operators, is only possible if $c = 0$. This holds for
any choice of $f$ within the above limitations, so 
$V(\lambda)$ acts trivially on ${\cal R}(W)$. Taking also into 
account that $\Omega$ is cyclic for ${\cal R}(W)$, one
arrives at $V(\lambda) = 1$, $\lambda \in \mathbb R$,
from which the first part of statement $(i)$ follows. 

Similarly, modular
theory and the theorem of Borchers mentioned above
imply that the unitary operator $I \doteq J_W J$
commutes with all Poincar\'e transformations $U(x,B)$
and, taking into account relation (\ref{wedgelocality}),
one also has 
$I {\cal R}(W) I^{-1} \subset {\cal R}(W)$.
Hence, putting $\phi_I(f) \doteq I \phi(f) I^{-1}$,
one finds by the same reasoning as in the preceding step
that $\phi_I(f) = \phi(f)$. Thus $I=1$,
proving the second part of statement $(i)$. The
statement about Haag duality then follows from the equalities
\begin{equation}
 {\cal R}(W)^\prime = J_W  {\cal R}(W) J_W
= J  {\cal R}(W) J =  {\cal R}(-W).
\end{equation}

It remains to prove the assertion that ${\cal R} (W)^\prime \Omega $
is a core for the selfadjoint operators $\phi(f)$, $\phi_\lambda(f)$
and $\phi_I(f)$, respectively. To this end 
one makes use of bounds, given in  \cite{Le},
on the action of the field operators on $n$--particle states $\Psi_n$.
One has $\| \phi(f) \Psi_n \| \leq c_f (n +1)^{1/2} \|  \Psi_n \|$,
where $c_f$ is some constant depending only on $f$.
Since the field operators change the particle number at most
by $\pm 1$, one can proceed from this estimate to corresponding 
bounds for $\Psi \in\cal D$,  given by 
$\| \phi(f) \Psi \| \leq 2 c_f \, \| (N + 1)^{1/2} \Psi \|$,
where $N$ is the particle number operator. Recalling that 
$P_0$ denotes the (positive) generator of the time translations,
it is also clear that $m \, (N + 1) \leq  (P_0 + m1)$. So 
for $\Psi \in {\cal D} \cap {\cal D}_0$,
where ${\cal D}_0$ is the domain of  $P_0$, one arrives at the inequalities
\begin{equation}
\| \phi(f) \Psi \| \leq 2 c_f \, \| (N + 1)^{1/2} \Psi \|
\leq 2 m^{-1/2} c_f \, \| (P_0 + m1)^{1/2} \Psi \|.
\end{equation}
It follows from this estimate by standard arguments that any core for
$P_0$ is also a core for the field operators $\phi(f)$.
Since the unitary operators $V(\lambda)$ and $I$ in the 
preceding steps were shown to commute with the time translations,
this domain property is also shared by the transformed field operators 
$\phi_\lambda (f)$ and $\phi_I (f)$, respectively. 

In order to complete the proof, one has only to show that  
${\cal R} (W)^\prime \Omega \cap {\cal D}_0 $ is a core for $P_0$.
Now ${\cal R} (W)^\prime \Omega$
is mapped into itself by all translations $U(x)$, $x \in -W$.
Hence,  taking into account the invariance of $\Omega$
under translations, one finds that 
$\widetilde{f}(P) {\cal R} (W)^\prime \Omega
\subset {\cal R} (W)^\prime \Omega \cap {\cal D}_0$
for any test function $f$ with $\text{supp} f \subset - W$. 
But this space of functions contains elements $f$ such 
that $ \widetilde{f}(P)$ is invertible. Hence 
$(P_0 \pm i 1) \widetilde{f}(P) {\cal R} (W)^\prime \Omega 
\subset (P_0 \pm i 1) ({\cal R} (W)^\prime \Omega \cap {\cal D}_0)$ 
both are dense subspaces of $\cal H$, proving the statement.
\end{proof}

In view of the covariance properties of the net, it is
apparent that analogous statements hold for all wedge algebras.
Thus the only point left open in this reconstruction of a 
relativistic quantum field theory from scattering data
is the question of whether the wedge 
algebras contain operators which can be interpreted as observables
localized in finite spacetime regions,
such as the double cones $C_{x,y} \doteq (W + y) \cap (-W + x)$,
$x - y \in W$. By Einstein causality, observables localized in $C_{x,y}$
have to commute with all operators localized in the adjacent 
wedges $W + x$ and $-W + y$. They are therefore elements of 
the  algebra
\begin{equation}
{\cal R}(C_{x,y}) \doteq {\cal R} (W + x)^\prime \cap  
{\cal R} (-W + y)^\prime =
{\cal R} (-W + x) \cap  {\cal R} (W + y).
\end{equation} 
It follows from the properties of the wedge algebras 
established thus far that the resulting 
map ${\cal C} \mapsto {\cal R}(C)$ from double cones to von 
Neumann algebras defines a local and Poincar\'e covariant net
on Minkowski space. So if the theory describes local 
observables, the algebras $ {\cal R}(C)$ are to be non--trivial.

At this point the nuclearity condition formulated in
Sec.\ 2 comes in. Knowing by the preceding 
proposition the explicit form of the modular operator 
affiliated with $({\cal R} (W), \Omega)$
and taking into account the invariance of $\Omega$
under spacetime translations, we are led to consider,
for given  $x \in W$, the maps 
\begin{equation} \label{qftnuclearity}
A \mapsto  U(0,B(-i \pi / 2)) \, U(x,1) \, A \Omega, \quad A \in {\cal R} (W).
\end{equation}
Within the present context one then has the following 
more concrete version of Proposition \ref{prop}.
\begin{proposition} Let the maps (\ref{qftnuclearity}) be nuclear, 
$x \in W$. Then \\[1mm]
(i) the net of wedge algebras has the split property; \\[1mm]
(ii) for any open double cone 
$C \subset {\mathbb R}^2$ the corresponding algebra ${\cal R}(C)$
is isomorphic to the unique hyperfinite factor of type III$_1$.
In particular it has cyclic and separating vectors.
\end{proposition}
So in order to establish the existence of
local operators in the theory, one needs an estimate  
of the size of the set of vectors 
\begin{equation} \label{set}
U(0,B(-i \pi / 2)) \, U(x,1) \, A \Omega,
\quad A \in  {\cal R} (W)_1,
\end{equation}
\ie the image of the unit ball $ {\cal R} (W)_1$ 
under the action of the map (\ref{qftnuclearity}).
We briefly indicate here the steps required in such an analysis
which are similar to those carried out in  \cite{Br} in 
an investigation of the Haag--Swieca compactness condition;
a more detailed account of these results will be presented 
elsewhere.

Making use of the localization properties of the operators 
$ A \in  {\cal R} (W)$ and the analyticity properties of
the scattering function $S_2$, one can show that the 
$n$--particle wave functions
\begin{equation}
\theta_1, \dots \theta_n \mapsto 
(A \Omega)_n (\theta_1, \dots \theta_n) 
\end{equation} 
extend, in the sense of distributions, to analytic functions in the domain
$ 0 < \text{\rm Im} \, \theta_i < \pi$, 
$- \delta < \text{\rm Im} \, (\theta_i - \theta_k)  < \delta$, where 
$i,k = 1, \dots n$ and $\delta$ depends on the domain of 
analyticity of the scattering function $S_2$. Thus the functions
\begin{equation} \label{normal}
\begin{split} 
\theta_1, \dots & \theta_n \mapsto ( U(0,B(-i\pi /2)) \,A \Omega)_n 
(\theta_1, \dots \theta_n) \\ 
& = (A \Omega)_n (\theta_1 + i\pi/2, \dots \theta_n + i\pi/2) 
\end{split}
\end{equation} 
are analytic in the domain 
$-\delta/n < \text{\rm Im} \, \theta_i < \delta/n$, \ 
$i = 1, \dots n$. As a matter of fact, if $A \in {\cal R}(W)_1$,
the family of these functions turns out to be
uniformly bounded (normal) on this domain.
Taking also into account that $U$ is a representation of 
the Poincar\'e group, one obtains for $x \in W$ the equality
\begin{equation} \label{grouprelation}
U(0,B(-i \pi / 2)) \, U(x,1) \, A \Omega =
e^{x_1 P_0 - x_0 P_1} \, U(0,B(-i \pi / 2)) \, A \Omega,
\end{equation}
so the $n$--particle components of the vectors  (\ref{set})
have wave functions of the form
\begin{equation} \label{wavefunctions}
\begin{split}
\theta_1, & \dots \theta_n \mapsto
( U(0,B(-i\pi /2)) \, U(x,1) \,A \Omega)_n 
(\theta_1, \dots \theta_n) \\
& = e^{\mbox{\footnotesize
$m \sum_{k = 1}^n (x_1\, \text{\rm ch} (\theta_k) 
- x_0 \, \text{\rm sh} (\theta_k))$}} \,
 (A \Omega)_n (\theta_1 + i\pi/2, \dots 
\theta_n + i\pi/2). 
\end{split}
\end{equation}
Since, for $x \in W$, the exponential factor gives rise to 
a strong damping of large rapidities,  
it follows from the preceding results
that the wave functions (\ref{wavefunctions}) 
form, for $A \in {\cal R}(W)_1$,
a bounded subset of the space of test functions 
${\cal S} ({\mathbb R}^n)$ and hence a  
nuclear subset of 
$ {L}^2({\mathbb R}^n) = {\cal H}_n $.
Moreover, taking into account the spectral properties 
of $(P_0,P_1)$, 
relation (\ref{grouprelation}) combined with
the estimate 
$\|  U(0,B(-i\pi /2)) \,A \Omega \| \leq \| A \|$ 
following from modular theory implies 
\begin{equation} 
\| ( U(0,B(-i\pi /2)) \, U(x,1) \,A \Omega)_n  \| \leq
e^{n \, m (x_1 + |x_0|) }, \quad A \in {\cal R}(W)_1. 
\end{equation}
So these norms tend rapidly to $0$ for large $n \in {\mathbb N}$
if $x \in W$.
Combining these facts, one finds after a moments 
reflection that the sets (\ref{set}) are relatively compact
in ${\cal H}$, implying that the maps  (\ref{qftnuclearity})
are compact. So they can be approximated with arbitrary
precision by finite sums of maps of rank one. 
In order to prove that they are also nuclear, 
one needs more refined estimates, however.

\section{An instructive example}
\setcounter{equation}{0}
In order to illustrate the quantitative estimates
needed for the proof that the map (\ref{qftnuclearity})
is nuclear, we consider here the case of 
trivial scattering, $S_2 = 1$, \ie the theory 
of a free massive Bose field $\phi$. There the combinatorial 
problems appearing in the analysis of the size 
of sets of the type (\ref{set}) have been settled 
in \cite{BuJa} and we shall make use of these results here.

We begin by recalling some well known facts:
The restrictions of the field $\phi$
and of its time derivative  $\dot{\phi}$ to the 
time zero plane are operator valued distributions
on the domain $\cal D$. These time zero fields,  
commonly denoted by $\varphi$ and $\pi$, 
satisfy canonical equal time commutation relations. 
If smeared with test functions
$h$ having support in the interval $(-\infty,0)$, they generate the 
von Neumann algebra $ {\cal R} (W)$ and, applying them to 
the vacuum vector $\Omega$, they create closed subspaces 
${\cal L}_\varphi(W)$, ${\cal L}_\pi(W)$
of the single particle space ${\cal H}_1$ given by  
\begin{equation}
\begin{split}
& {\cal L}_\varphi(W) = \{ \theta \mapsto \widetilde{h} (m \, 
\text{sh}(\theta)) : \text{supp} \,  h \subset (-\infty,0) \}^-, \\
& {\cal L}_\pi(W) = \{ \theta \mapsto \text{ch}(\theta) \, \widetilde{h} (m \, 
\text{sh}(\theta)) : \text{supp} \,  h \subset (-\infty,0) \}^-,
\end{split}
\end{equation}
where the tilde denotes Fourier transformation. We also consider 
the shifted 
subspaces $ {\cal L}_\varphi(W+x) \doteq U(x,1) \,  {\cal L}_\varphi(W)$ and
$ {\cal L}_\pi(W+x) \doteq U(x,1) \,  {\cal L}_\pi(W)$
and denote the corresponding orthogonal projections by
$E_\varphi(W+x)$ and $E_\pi(W+x)$, respectively. After these
preparations we are in a position to apply  
the results in \cite[Thm.\ 2.1]{BuJa}
which we recall here for the convenience of the reader in a form
appropriate for the present investigation.
\begin{lemma} Consider the theory with scattering 
function $S_2 = 1$ and let 
$E_\varphi(W+x) \, U(0,B(-i\pi/2))$
and $E_\pi(W+x) \, U(0,B(-i\pi/2))$ be 
trace class operators with operator norms less than 1,
$x \in W$.  Then the sets (\ref{set}) are nuclear.
\end{lemma}
\noindent Thus 
the proof that the modular nuclearity condition 
is satisfied in the present theory reduces to a problem of spectral analysis 
in the single particle space ${\cal H}_1$. 
We first turn to the task of providing estimates of the norms of the 
operators appearing in the lemma.

Let $\Phi_h \in {\cal L}_\varphi(W)$ be a vector with wave function
$\theta \mapsto \Phi_h(\theta) =  \widetilde{h} (m \, 
\text{sh}(\theta))$, where $h$ is, as before, a test function with
support in $(-\infty,0)$. Because of these support properties,
$\Phi_h$ lies in the domain of all boosts 
$ U(0,B(\theta))$ for complex $\theta$
with $-\pi \leq \text{Im} \, \theta \leq 0$.
Furthermore, as $\text{sh}(\theta + i \pi) = \text{sh}(-\theta)$, one has 
$(U(0,B(-i\pi)) \Phi_h)(\theta) 
= \widetilde{h} (m \, \text{sh}(-\theta)) = \Phi_h (-\theta)$.
But this implies  $\|  U(0,B(-i\pi)) \,  \Phi_h \| = \|  \Phi_h \|$
and consequently $\|  U(0,B(-i\pi/2)) \,  \Phi_h \| \leq \|  \Phi_h \|$.
Making use now of the 
properties of the representation $U$, one obtains 
the estimate, $x \in W$, 
\begin{equation}
\begin{split}
&  \|  U(0,B(-i\pi/2))  \, U(x,1) \,  \Phi_h \| 
 = \| e^{x_1 P_0 - x_0 P_1} \, U(0,B(-i \pi / 2)) \,  \Phi_h \| \\
& \leq e^{m(x_1 + |x_0|)} \,  \|  U(0,B(-i\pi/2)) \, \Phi_h \| 
  \leq  e^{m(x_1 + |x_0|)} \,  \| U(x,1) \,  \Phi_h \|. 
\end{split}
\end{equation}
Since $ \Phi_h$ was arbitrary within the above limitations
and $(x_1 + |x_0|)$ is negative, this yields the norm estimate
$\| U(0,B(-i\pi/2)) \, E_\varphi (W+x)\| < 1$,  $x \in W$. But 
the adjoint operator $ E_\varphi (W+x)  \, U(0,B(-i\pi/2))$
has the same norm, so the desired bound follows. In a similar  manner
one can show that the operator $ E_\pi (W+x)  \, U(0,B(-i\pi/2))$
also has norm less than $1$.

It remains to establish the trace class property of these
operators. To this end we consider the restriction of the 
operator $ U(0,B(-i\pi/2)) \, U(x,1) $, $x \in W$, to the
subspaces ${\cal L}_\varphi(W)$ and  ${\cal L}_\pi(W)$,
respectively. Let, as before,  $\Phi_h \in {\cal L}_\varphi(W)$, then
\begin{equation}
( U(0,B(-i\pi/2))  \, U(x,1) \,  \Phi_h)(\theta) =
e^{\mbox{\footnotesize $x_1p_0(\theta) - x_0 p_1(\theta)$}} 
\, \Phi_h(\theta + i\pi/2).
\end{equation}
Making use of the analyticity and boundedness 
properties of $\theta \mapsto \Phi_h(\theta)$
and the fact that $\Phi_h(\theta + i\pi) = \Phi_h(-\theta)$,
one can represent  $\Phi_h(\theta + i\pi/2)$ by a Cauchy integral, 
\begin{equation} \label{f1}
 \Phi_h(\theta + i\pi/2) = \frac{1}{2 \pi i} \,
\int \! d\theta^\prime \, \Big\{ \frac{1}{\theta^\prime - \theta -i\pi/2}
+ \frac{1}{\theta^\prime + \theta -i\pi/2} \Big\} \, \Phi_h(\theta^\prime).
\end{equation}
Next, for $x \in W$, let $X_\varphi$ be the operator on ${\cal H}_1$ with 
kernel
\begin{equation} \label{f2} 
X_\varphi (\theta, \theta^\prime) = \frac{1}{2 \pi i} \,
e^{\mbox{\footnotesize $x_1p_0(\theta) - x_0 p_1(\theta)$}} \,
\Big\{ \frac{1}{\theta^\prime - \theta -i\pi/2}
+ \frac{1}{\theta^\prime + \theta -i\pi/2} \Big\}.
\end{equation}
Being the sum of products of 
multiplication operators 
in rapidity space, respectively its dual space,
which are bounded and rapidly decreasing,  
it is apparent that  $X_\varphi$ is of trace class.
Moreover, by the preceding results, 
$U(0,B(-i\pi/2)) \, E_\varphi(W+x) = X_\varphi \,  E_\varphi(W) \,
 U(x,1)^{-1}$. Since the trace class operators form a 
*--ideal in ${\cal B}({\cal H}_1)$, if follows that 
$U(0,B(-i\pi/2)) \, E_\varphi(W+x)$ and 
its adjoint $ E_\varphi(W+x) \, U(0,B(-i\pi/2))$ are of trace class.

By a similar argument one can also establish the trace class property
of $ E_\pi(W+x) \, U(0,B(-i\pi/2))$, the only difference being
that for  vectors $\Phi_h \in {\cal L}_\pi(W)$  with wave functions
$\theta \mapsto \Phi_h(\theta) = \text{ch}(\theta) \, \widetilde{h} (m \, 
\text{sh}(\theta))$ one now has $\Phi_h(\theta + i \pi)
= - \Phi_h(-\theta)$. As a consequence, the sum in relation
(\ref{f1}) turns into a difference, but this does not affect the  
conclusions. So the following statement has been proven.
\begin{proposition} In the theory with scattering function
$S_2 = 1$, the sets (\ref{set}) and corresponding maps
\begin{equation}  
A \mapsto  U(0,B(-i \pi / 2)) \, U(x,1) \, A \Omega, \quad A \in {\cal R} (W).
\end{equation}
are nuclear, $x \in W$.
\end{proposition}
We thus have verified in the present model 
the modular nuclearity condition for wedge algebras
with all of its consequences. In particular, 
the wedge algebras have the split property. Although 
the latter fact was known before \cite{Mue},
there did not yet exist a proof 
in the literature. 

By similar arguments one can also 
treat the theory with scattering function $S_2 = -1$, \ie 
the theory of a (non--local) free massive Fermi field. There 
one expects that the sets (\ref{set}) are somewhat smaller 
than in the present case because of the Pauli principle. 
It is even more challenging, however, to provide a quantitative estimate 
of the size of the sets (\ref{set}) in theories with
generic scattering function. This problem will be tackled elsewhere.

\section{Conclusions}
\setcounter{equation}{0}

Within the algebraic setting of quantum field 
theory, we have presented a method which allows one 
to decide whether algebras affiliated with wedge shaped
regions in two--dimensional Minkowski space contain 
compactly localized operators. This method seems to  
be particularly useful for proving the existence of local
operators in theories with factorizing $S$--matrix.
It is thus complementary to the formfactor program,
where one tries to exhibit such operators explicitly 
by solving an infinite system of equations.

The upshot of the present investigation is the insight 
that the basic algebraic problem of checking locality, 
which amounts to computing relative commutants, can 
be replaced by an analysis of spectral properties 
of representations of the Poincar\'e group. 
There exist other methods by which the 
crucial intermediate step in our argument, the proof of the 
split property of wedge algebras, could be accomplished 
\cite{Bu,BuDaLo2,DaLo,BuWi,DaFr,Mue1}. But the present 
approach requires less {\it a priori} information about the 
underlying theory and also seems better managable in
concrete applications. Moreover, in view of the fact
that it relies only on the modular structure, it is applicable 
to theories on arbitrary spacetime manifolds. 

It is apparent, however, that the split property of 
wedge algebras is in general an unnecessarily strong requirement if one is 
merely interested in the existence of compactly localized
operators. As a matter of fact, it follows from an argument of Araki
that it cannot hold in more than two spacetime dimensions, cf. 
\cite[Sec.\ 2]{Bu}. It would therefore be desirable to
establish less stringent conditions which still imply that 
the relative commutant fixed by
a given inclusion of von Neumann algebras is non--trivial.
The present results seem to suggest that this information is 
encoded in spectral properties of the corresponding modular 
operators, but a clarification of 
this point requires some further analysis.

An appropriately weakened condition which would 
allow one to establish the existence of local operators  
in non--local algebras also in higher dimensions 
would have several interesting applications. 
This existence problem was recently met
in the context of theories of massless 
particles with infinite spin \cite{MuSchYn}, for example. It 
also appears in the algebraic approach to the construction of theories 
of particles with anyonic statistics \cite{Mu2}  
and the construction of nets of wedge algebras from
information on the modular data \cite{Bo3,KaWi,Wi1,Wi2,BuDrFlSu}. 
A solution of this problem would thus be a major step
in the algebraic approach to constructive 
problems in local quantum physics.

\vspace*{8mm}
\noindent {\Large \bf Acknowledgements} \\[3mm]
We would like to thank the Deutsche Forschungsgemeinschaft
for financial support.

\end{document}